\documentclass[smallcondensed]{svjour3}     
\smartqed  
\usepackage{graphicx}
\usepackage{amssymb}
\usepackage{graphicx}

\newcommand{\D}{\mbox{$D$}}                 
\newcommand{\DD}{\mbox{$\mathcal{D}$}}      
\newcommand{\F}{\mbox{$F$}}                 

\newcommand{\G}[2]{\mbox{$\mathcal{G}_{#1,#2}$}}
\newcommand{\CH}{\mbox{$\G{D}{F}$}}                 
\newcommand{\rep}[2]{\mbox{$repairs(#1,#2)$}}       
\newcommand{\MM}[1]{\mbox{$\mathcal{MM}(#1)$}}      
\newcommand{\edges}[3]{\mbox{$edges_{#1,#2}(#3)$}}  
\newcommand{\SMM}{\mbox{$\mathcal{S}$}}           
\newcommand{\Set}{\mbox{$\mathcal{R}$}}

\newcommand{\Dfr}{\mbox{$\DD(D,F)$}} 

\newcommand{\Dmi}{\mbox{$\mathcal{D}_{min}$}}
\newcommand{\Dmin}[2]{\mbox{$\Dmi(#1,#2)$}}
\newcommand{\Cl}{\mbox{$C$}} 
\newcommand{\Gr}{\mbox{$G$}} 
\newcommand{\cliques}[2]{\mbox{$cliques(#1,#2)$}}
\newcommand{\cls}{\mbox{$clusters$}}
\newcommand{\reduced}{\mbox{$reduction$}}
\newcommand{\canonical}{\mbox{canonical}}
\newcommand{\repair}{\mbox{S-repair}}

\def\<{\mbox{$\langle$}}
\def\>{\mbox{$\rangle$}}

\begin{document}

\title{Disjunctive Databases for Representing Repairs
}


\author{Cristian Molinaro \and
        Jan Chomicki \and
        Jerzy Marcinkowski
}


\institute{
Cristian Molinaro \at
DEIS, Universit\'a della Calabria, 87036 Rende, Italy \\
\email{cmolinaro@deis.unical.it}
\and
Jan Chomicki \at
Department of Computer Science and Engineering, 201 Bell Hall, The State University of New York at Buffalo,
Buffalo, NY 14260, USA \\
\email{chomicki@cse.buffalo.edu}
\and
Jerzy Marcinkowski \at
Institute of Informatics, Wroclaw University, Przesmyckiego 20, 51-151 Wroclaw, Poland\\
\email{jma@cs.uni.wroc.pl}
}

\date{Received: date / Accepted: date}

\maketitle

\begin{abstract}
This paper addresses the problem of representing the set of repairs
of a possibly inconsistent database by means of a disjunctive database.
Specifically, the class of denial constraints is considered.
We show that, given a database and a set of denial constraints, there exists a (unique)
disjunctive database, called \emph{\canonical}, which represents the repairs of the database w.r.t. the constraints
and is contained in any other disjunctive database with the same set of minimal models.
We propose an algorithm for computing the canonical disjunctive database.
Finally, we study the size of the canonical disjunctive database in the presence of functional
dependencies for both repairs and cardinality-based repairs.
\end{abstract}

\section{Introduction}

The problem of managing inconsistent data nowadays arises in several scenarios.
How to extract reliable information from {\em inconsistent databases}, i.e. databases violating integrity constraints,
has been extensively studied in the past several years.
Most of the works in the literature rely on the notions of \emph{repair} and
\emph{consistent query answer}~\cite{DBLP:conf/pods/ArenasBC99}.
Intuitively, a repair for a database w.r.t. a set of integrity constraints is a consistent database
which ``minimally'' differs from the (possibly inconsistent) original database.
The consistent answers to a query over an inconsistent database are those tuples which can be
obtained by evaluating the query in every repair of the  database.
Let us illustrate the notions of repair and consistent query answer by means of an example.

\begin{example}\label{ex:intro}
Consider the following relation $r$
\begin{center}
\begin{tabular}{|c|c|c|}
\multicolumn{3}{c} {$ employee$}\\
\hline
$Name$   &  $Salary$  &   $Dept$\\
\hline
$john$   &  $50$      &   $cs$\\
$john$   &  $100$     &   $cs$\\
\hline
\end{tabular}
\end{center}

\noindent
\\ and the functional dependency $f:\ Name \rightarrow Salary\ Dept$ stating that each employee has a unique salary
and a unique department.
Clearly, $r$ is inconsistent w.r.t. $f$ as it stores two different salaries for the same employee $john$.
Assuming that the database is viewed as a set of facts and the symmetric difference is used to capture the distance between two databases,
there exist two repairs for $r$ w.r.t. $f$, namely $\{employee(john,50,cs)\}$
and $\{employee(john,100,cs)\}$.
The consistent answer to the query asking for the department of $john$ is $cs$
(as this is the answer of the query in both  repairs),
whereas the query asking for the salary of $john$ has no consistent answer
(as the two repairs do not agree on the answer).
\end{example}

An introduction to the central concepts of consistent query answering is~\cite{DBLP:conf/icdt/Chomicki07},
whereas surveys on this topic are~\cite{DBLP:conf/dagstuhl/BertossiC03,DBLP:journals/sigmod/Bertossi06}.

Inconsistency leads to {\em uncertainty} as to the actual values of tuple attributes.
Thus, it is natural to study the possible use of incomplete database frameworks in this context.
The set of repairs for a possibly inconsistent database could be represented by means of an incomplete database
whose possible worlds are exactly the repairs of the inconsistent database.

In this paper, we consider a specific incomplete database framework: {\em disjunctive databases}.
A disjunctive  database is a finite set of disjunctions of facts.
Its semantics is given by the set of minimal models.
There is a clear intuitive connection between inconsistent and disjunctive databases.
For instance, the repairs of the relation $r$ of Example~\ref{ex:intro} could be represented by the disjunctive database
$\DD=\{employee(john,50,cs) \vee employee(john,100,cs)\}$, as the minimal models of $\DD$ are exactly the repairs
of $r$ w.r.t. $f$. Disjunctive databases have been studied for a long time
\cite{ImNaVa91,ImVMVa95,DBLP:conf/birthday/MinkerS02,DBLP:conf/icdt/FernandezM92}. More recently,
they have again attracted attention in the database research community because of potential applications in data
integration, extraction and cleaning \cite{BSHTW08}.
Our approach should be distinguished from the approaches that rely on stable model semantics of {\em disjunctive logic programs with negation} to represent
repairs of inconsistent databases \cite{ArBeCh03,CaLeRoIJCAI03,GrGrZu03}.

In this paper we address the problem of {\em representing} the set of repairs of a database
w.r.t. a set of denial constraints by means of a disjunctive database (in other words,
a disjunctive database whose minimal models are the repairs).

We show that, given a database and a set of denial constraints, there exists a unique, {\em canonical}
disjunctive database which (a) represents the repairs of the database w.r.t. the constraints, and (b) is
contained in any other disjunctive database having the same set of minimal models.
We propose an algorithm for computing the canonical disjunctive database  which in general can be of exponential size.
Next, we study the size of the canonical disjunctive database in the presence of restricted functional dependencies.
We show that the canonical disjunctive database is of linear size when only one key in considered,
but it may be of exponential size in the presence of two keys or one non-key functional dependency.
Finally, we demonstrate that these results hold also for a different, cardinality-based semantics of repairs~\cite{LoBe07}.

The paper is organized as follows.
In Section~\ref{sec:preliminaries}, we introduce some basic notions in inconsistent and disjunctive databases.
In Section~\ref{sec:algo+theo}, we present an algorithm to compute the canonical disjunctive database
and show that this database is contained in any other disjunctive database with the same minimal models.
In Section~\ref{sec:specialcases}, we study the size of the canonical disjunctive databases in
the presence of functional dependencies.
In Section~\ref{sec:card-repairs}, we investigate the size of the canonical disjunctive databases under the
cardinality-based semantics of repairs.
Finally, in Section~\ref{sec:conclusions} we draw the conclusions and outline some possible future research topics.

\section{Preliminaries}\label{sec:preliminaries}
In this section we introduce some basic notions of relational, inconsistent, and disjunctive databases.

\subsection{Relational databases}

We assume the standard concepts of the relational data model. A database is a collection of relations.
Each relation is a finite set of tuples and has a finite set of attributes. The values
of each attribute are integers, rationals or uninterpreted constants.
Each tuple $\bar{t}$ in a relation $p$ can be viewed as a fact $p(\bar{t})$;
then a database can be viewed as a finite set of facts.

We say that a database is \emph{consistent} w.r.t. a set of integrity constraints if
it satisfies the integrity constraints, otherwise it is \emph{inconsistent}.
In this paper we consider the class of \emph{denial constraints}.
A denial constraint is a first-order logic sentence of the following form:
\[
\forall \overline{X}_1 \dots \overline{X}_n\ \neg [p_1(\overline{X}_1) \wedge \dots \wedge p_n(\overline{X}_n)
\wedge \varphi(\overline{X}_1, \dots, \overline{X}_n)]
\]
\noindent
where the $\overline{X}_i$'s are sequences of variables, the $p_i$'s are relational symbols and
$\varphi$ is a conjunction of atoms referring to built-in, arithmetic or comparison, predicates.
Special cases of denial constraints are functional dependencies and key constraints.
A functional dependency is of the form
\[
\forall \overline{X}_1\overline{X}_2\overline{X}_3\overline{X}_4\overline{X}_5\ \neg
[p(\overline{X}_1,\overline{X}_2,\overline{X}_4) \wedge p(\overline{X}_1,\overline{X}_3,\overline{X}_5)
\wedge \overline{X}_2 \neq \overline{X}_3]
\]
The previous functional dependency can be also stated as $X \rightarrow Y$,
where $X$ is the set of attributes of $p$ corresponding to $\overline{X}_1$ whereas
$Y$ is the set of attributes of $p$ corresponding to $\overline{X}_2$ (and $\overline{X}_3$).
A key constraint is of the form
\[
\forall \overline{X}_1\overline{X}_2\overline{X}_3\ \neg
[p(\overline{X}_1,\overline{X}_2) \wedge p(\overline{X}_1,\overline{X}_3)
\wedge \overline{X}_2 \neq \overline{X}_3]
\]
We say that the set of attributes corresponding to $\overline{X}_1$ is a key.
We assume that the given set of integrity constraints is satisfiable.

\subsection{Inconsistent databases}

As it has been already said in the introduction, a repair of a database  w.r.t. a set of integrity constraints
is a consistent database which ``minimally'' differs from the (possibly inconsistent) original database~\cite{DBLP:conf/pods/ArenasBC99}.
The symmetric difference is used to capture the distance between two databases.
Because we consider denial constraints and assume that the symmetric difference has to be minimal under set inclusion,
repairs are {\em maximal consistent subsets} of the original database
(although in Section~\ref{sec:card-repairs} we will consider cardinality-based repairs,
where the {\em cardinality} of the symmetric difference is minimized).
The set of repairs of a database $\D$ w.r.t. a set $\F$ of denial constraints
is denoted by $\rep{\D}{\F}$.

Given a database $\D$ and a set $\F$ of denial constraints, the \emph{conflict hypergraph}~\cite{DBLP:journals/iandc/ChomickiM05}
for $\D$ and $\F$,
denoted by \CH, is a hypergraph whose set of vertices is the set of facts of $\D$,
whereas the set of edges consists of all the sets $\{p_1(\overline{c}_1), \dots, p_n(\overline{c}_n)\}$ s.t.
$p_1(\overline{c}_1), \dots, p_n(\overline{c}_n)$ are facts of $\D$ which violate together a denial constraint in \F, i.e.
there exist a denial constraint
\[
\forall \overline{X}_1 \dots \overline{X}_n\ \neg [p_1(\overline{X}_1) \wedge \dots \wedge p_n(\overline{X}_n)
\wedge \varphi(\overline{X}_1, \dots, \overline{X}_n)]
\]
in $F$ and a substitution $\rho$ s.t. $\rho(\overline{X}_i)=\overline{c}_i$ for $i=1..n$ and
$\varphi(\overline{c}_1,\dots,\overline{c}_n)$ is true.
A fact $t$ of $\D$ is said to be \emph{conflicting} (w.r.t. $\F$)
if it is involved in some constraint violations, that is there exists
an edge $\{t,t_1,\dots,t_m\}\ (m \geq 0)$ in \CH.
For a fact $t$ of $\D$, we denote by $\edges{D}{F}{t}$ the set
of edges of $\CH$ containing $t$, i.e. $\edges{D}{F}{t} = \{ e= \{t,t_1,\dots,t_k\}\ |\ e \in E\}$.

\subsection{Disjunctive databases}
A disjunctive database $\DD$ is a finite set of non-empty disjunctions of distinct facts.
A disjunction containing exactly one fact is called a \emph{singleton} disjunction.
A set $M$ of facts is a model of $\DD$ if $M \models \DD$; $M$ is minimal
if there is no $M' \subset M$ s.t. $M'\models \DD$.
We denote by $\MM{\DD}$ the set of minimal models of $\DD$.
For a disjunction $d \in \DD$, $S_d$ denotes the set of facts appearing in $d$.
Given two distinct disjunctions $d_1$ and $d_2$ in $\DD$, we say that $d_1$ \emph{subsumes} $d_2$
if the set of facts appearing in $d_1$ is a (proper) subset of the set of facts appearing in $d_2$,
i.e. $S_{d_1} \subset S_{d_2}$.
Moreover, the reduction of $\DD$, denoted by $\reduced(\DD)$, is the disjunctive database obtained from $\DD$ by discarding
all the subsumed disjunctions, that is
\[\reduced(\DD) = \{d\ |\ d \in \DD\ \wedge \nexists d' \in \DD\ \mbox{s.t.}\ d' \mbox{ subsumes } d\}.\]
Observe that for any disjunctive database $\DD$, $\MM{\DD}=\MM{\reduced(\DD)}$.

\subsection{Computational complexity}
We adopt here the \emph{data complexity} assumption \cite{Var82}, under which the complexity is a function
of the number of facts in the database. The set of integrity constraints is considered fixed.
In this setting, the conflict hypergraph is of polynomial size and can be computed in polynomial time.
We study the size of a disjunctive database representing the set of repairs of a relational database $\D$ w.r.t.
a set of integrity constraints $\F$ as a function of the number of facts in $\D$.

\section{Disjunctive databases for representing repairs}\label{sec:algo+theo}

In this section we propose an algorithm to compute a disjunctive database
whose minimal models are the repairs of a given database w.r.t. a set of denial constraints.
We show that the so computed disjunctive database is the canonical one,
that is any other disjunctive database whose minimal models coincide with the repairs of the original database
is a superset of the canonical one (containing, in addition, only disjunctions which are subsumed by
disjunctions in the canonical disjunctive database).


\begin{figure}[h]\label{fig:algorithm1}
{\small
\hrule
$ $\\
\noindent
\textbf{Algorithm 1}\\
\textbf{Input:} a database \D\ and a set \F\ of denial constraints\\
\textbf{Output:} a disjunctive database whose minimal models are the repairs for \D\ and \F\\
\[
\begin{array}{rll}
1:  & & \mbox{$\widehat{\DD}:=\emptyset$}\\
2:  & & \mbox{$\D':= D -\ \{t\ |\ \{t\}$ is an edge of $\CH \}$}\\
3:  & & \mbox{for each $t \in \D'$}\\
4:  & & \mbox{\hspace*{4mm} Let $\edges{D'}{F}{t}=\{e_1,\dots,e_n\}$}\\
5:  & & \mbox{\hspace*{4mm} $\widehat{\DD}:= \widehat{\DD} \cup \{t \vee t_1 \vee \dots \vee t_n\ |\ t_i \in e_i \mbox{ and } t_i\neq t$ for $i=1..n\}$}\\
6:  & & \mbox{repeat until $\widehat{\DD}$ does not change}\\
7:  & & \mbox{\hspace*{4mm}for each edge $e = \{t_1, \dots, t_k\}$ in $\mathcal{G}_{D',F}$}\\
8:  & & \mbox{\hspace*{8mm}for each $t_1 \vee D_1, \dots , t_k \vee D_k \in \widehat{\DD}$ s.t. $D_i$ is not an empty disjunction and}\\
    & & \mbox{\hspace*{8mm}$D_i$ does not contain any fact $t'\neq t_i$ in $e$, $i=1..\, k$}\\
9:  & & \mbox{\hspace*{13mm}$\widehat{\DD}:= \widehat{\DD} \cup \{D_1 \vee \dots \vee D_k\}$}\\
10: & & \mbox{return $\reduced(\widehat{\DD})$}\\
\end{array}
\]
\hrule $ $\\
}
\end{figure}

We denote by \Dfr\ the disjunctive database returned by Algorithm~1
with the input consisting of a database $\D$ and a set $\F$ of denial constraints.
In the second step of the algorithm, every fact $t$ s.t. $\{t\}$ is an edge of the conflict hypergraph is discarded.

The disjunctions introduced in the step~5 allow us to guarantee that
the minimal models are maximal (consistent) subsets of $\D$.
Intuitively, a disjunction of the form $t \vee t_1 \vee \dots \vee t_n$ (which contains one fact
from each edge containing $t$) prevents from having a model $m$ of $\widehat{\DD}$ which contains
neither $t$ nor the $t_i$'s as in this case $m$ would not be maximal.

The disjunctions introduced in the step~9 allow us to guarantee that the minimal models of \Dfr\
are consistent w.r.t. $\F$.
Specifically, the loop in lines~6--9 is performed until $\widehat{\DD}$ satisfies the following property:
for every edge $e = \{t_1, \dots, t_k\}$ of the conflict hypergraph ($k>1$),
if there are $t_1 \vee D_1, \dots , t_k \vee D_k \in \widehat{\DD}$ s.t. each $D_i$ is not an empty disjunction,
then $\{D_1 \vee \dots \vee D_k\}$ is also in $\widehat{\DD}$.
As it is shown in the proof of Theorem~\ref{th:algo_correctness}, this property entails that every minimal
model of $\widehat{\DD}$ does not contain $\{t_1, \dots, t_k\}$.
Observe that the loop ends when $\widehat{\DD}$ does not change anymore;
at each iteration new disjunctions are added to $\widehat{\DD}$.
Since the number of disjunctions is bounded (if the original database has $h$ facts,
there cannot be more than $2^h-1$ disjunctions) the algorithm always terminates.
In the last step of the algorithm, subsumed disjunctions are deleted.
The following theorem states the correctness of Algorithm~1.

\begin{theorem}\label{th:algo_correctness}
Given a database $\D$ and a set $\F$ of denial constraints,
the set of minimal models of $\Dfr$ coincides with the set of repairs of $\D$ w.r.t. $\F$.\\

\em
\noindent \textbf{Proof.}
Since the the disjunctive database $\Dfr$ returned by Algorithm~1 is equal to $\reduced(\widehat{\DD})$
(step~10), then $\MM{\Dfr}=\MM{\widehat{\DD}}$.
First we prove \linebreak[4]
(1) $\rep{D}{F} \subseteq \MM{\widehat{\DD}}$ and next
(2) $\rep{D}{F} \supseteq \MM{\widehat{\DD}}$.

\noindent(1) Consider a repair $r$ in $\rep{D}{F}$.
First we show that (a) $r$ is a model of $\widehat{\DD}$ and next (b) that it is a minimal model.\\
(a) We prove that $r$ satisfies each disjunction in $\widehat{\DD}$ by induction.
Specifically, as base case we consider the disjunctions introduced
in the step~5 of the algorithm, whereas the inductive step refers to
the disjunctions introduced in the step~9.
Suppose by contradiction that $r$ does not satisfy
a disjunction $t \vee t_1 \vee \dots \vee t_n$ introduced in the step~5.
Observe that $\edges{D'}{F}{t} \subseteq \edges{D}{F}{t}$ and each edge $e'$ in
$\edges{D}{F}{t} - \edges{D'}{F}{t}$ is s.t. there is
a fact $t'\in e'$ s.t. $\{t'\}$ is an edge of \CH\ (clearly, $t'\not\in r$).
Since in each edge in $\edges{D}{F}{t}$ there is a fact (different from $t$)
which is not in $r$, then $r \cup \{t\}$ is consistent, which violates the maximality of $r$.
The inductive step consists in showing that $r$ satisfies any
disjunction added to $\widehat{\DD}$ in the step~9 assuming that $r$ satisfies $\widehat{\DD}$.
A disjunction $D_1 \vee \dots \vee D_k$, where the $D_i$'s  are not empty disjunctions,
is added to $\widehat{\DD}$ whenever there exist $t_1 \vee D_1, \dots , t_k \vee D_k$ in $\widehat{\DD}$
s.t. $e = \{t_1,\dots,t_k\}$ is an edge of $\mathcal{G}_{D',F}$,
and $D_i$ does not contain any fact $t'\neq t_i$ in $e$, for $i=1..k$.
Since $r$ satisfies all the disjunctions $t_1 \vee D_1, \dots , t_k \vee D_k$
and does not contain some fact $t_j$ in $e$ (as $e$ is an edge of \CH\ too),
it satisfies the disjunction $D_j$ and then $D_1 \vee \dots \vee D_k$ as well.
Hence $r$ is a model of $\widehat{\DD}$.\\
(b) We now show that $r$ is a minimal model, reasoning by contradiction.
Assume that there exists a model $m'\subset r$ and let $t$ be a fact in $r$ but not in $m'$.
Observe that $t$ is a conflicting fact (it cannot be the case that there is a model
of $\widehat{\DD}$ which does not contain a non-conflicting fact because the algorithm introduces, in the step~5,
a singleton disjunction $d$ for each non-conflicting fact $d$).
Moreover, as $r$ is a repair, $t$ is s.t. $\{t\}$ is not an edge of $\CH$ and then $t$ is in $D'$.
For each edge $e_i$ in $\edges{D'}{F}{t}=\{e_1,\dots,e_n\}$ there is
a fact $t_i \neq t$ which is not in $r$ as it is consistent and $\edges{D'}{F}{t}\subseteq \edges{D}{F}{t}$.
The same holds for $m'$ as it is a subset of $r$.
Then, the disjunction $t \vee t_1 \vee \dots \vee t_n$ in $\widehat{\DD}$ (added in the step~5)
is not satisfied by $m'$, which contradicts that $m'$ is a model.
Hence $r$ is a minimal model of $\widehat{\DD}$.\\

\noindent(2) Consider a minimal model $m$ in $\MM{\widehat{\DD}}$.
We show first (a) that it is consistent w.r.t. $F$ and then (b) that it is maximal.\\
(a) First of all, it is worth noting that $\widehat{\DD}$ doesn't contain a
singleton disjunction $t$ s.t. $t$ is a conflicting fact of $\D$.
This can be shown as follows.
Two cases may occur: either $\{t\}$ is an edge of $\CH$ or it is not.
As for the first case, since we have proved above that each repair
of $\D$ and $\F$ is a model of $\widehat{\DD}$ and no repair contains  $t$,
it cannot be the case that $t$ is a singleton disjunction of $\widehat{\DD}$.
Let us consider the second case.
For any conflicting fact $t$ in $\D$ s.t. $\{t\}$ is not an edge of $\CH$,
there exist a repair $r_1$ s.t. $t \in r_1$ and a repair $r_2$ s.t. $t \not\in r_2$.
As we have proved above, there are two minimal models of $\widehat{\DD}$
corresponding to $r_1$ and $r_2$, then it cannot be the case that
$t\in\widehat{\DD}$.
We prove that $m$ is consistent w.r.t. $F$ by
contradiction, assuming that $m$ contains a set of facts $t_1, \dots, t_k$
s.t. $e = \{t_1,\dots,t_k\}$ is in \CH.
Let $S_{t_i}=\{ D\ |\ t_i \vee D \in \widehat{\DD} \mbox{ and }
D\neq \emptyset \mbox{  does not contain any fact } t'\neq t_i \mbox{ in } e \}$
for $i=1..k$.
Two cases may occur: either (a) there is a set $S_{t_i}$ which is empty or
(b) all the sets $S_{t_i}$ are not empty.
(a) Let $t_j$ be a fact in $e$ s.t. $S_{t_j}$ is empty.
It is easy to see that $m - \{t_j\}$ is a model, which contradicts the minimality of $m$.
(b) For each $D_1 \in S_{t_1}, \dots, D_k \in S_{t_k}$,
it holds that $D_1 \vee \dots \vee D_k \in \widehat{\DD}$.
Then there is a set $S_{t_j}$ s.t. $m$ satisfies each $D$ in $S_{t_j}$,
otherwise it would be the case that some $D_1 \vee \dots \vee D_k$ in $\widehat{\DD}$,
where $D_i$ is in $S_{t_i}$ for $i=1..k$, is not satisfied.
It is easy to see that $m - \{t_j\}$ is a model, which contradicts the
minimality of $m$.
Hence $m$ is consistent w.r.t. $F$.\\
(b) Now we prove that $m$ is a maximal (consistent) subset of $\D$ reasoning by contradiction,
thus assuming that there exists $m' \supset m$ which is consistent.
Let $t$ be a fact in $m'$ but not in $m$.
Since $m'$ is consistent, for each edge $e_i$ in $\edges{D'}{F}{t}=\{e_1,\dots,e_n\}$ there is
a fact $t_i \neq t$ which is not in $m'$. 
The same holds for $m$ as it is a (proper) subset of $m'$.
This implies that $m$ doesn't satisfy the
disjunction $t \vee t_1 \vee \dots \vee t_n$ in $\widehat{\DD}$ (added in the step~5),
thus contradicting the fact the $m$ is a model.
Hence $m$ is a maximal consistent subset of $\D$, that is a repair.~\hfill$\Box$
\end{theorem}

Given a database $\D$ with $n$ facts, a rough bound on the
size of \Dfr\ is that it cannot have more than $2^n-1$ disjunctions and each disjunction contains at most $n$ facts,
for any set $\F$ of denial constraints
(in the next section we will study more precisely the size of \Dfr\ for special classes of denial constraints,
namely functional dependencies and key constraints).

The following theorem allows us to identify all the disjunctive databases which have
the same minimal models of a given disjunctive database.
Specifically, it states that given a disjunctive database $\DD$, any other disjunctive database with the same
minimal models is a superset of $\reduced(\DD)$ containing in addition only disjunctions subsumed
by disjunctions in $\reduced(\DD)$.
This result allows us to state that there is a (unique) disjunctive database
representing the repairs for a given database and a set of denial constraints which is contained
in any other disjunctive database with the same set of minimal models.
We call such a disjunctive database \emph{\canonical}.
Algorithm~1 computes the canonical disjunctive database (see Corollary~\ref{cor:algorithm-smallest}).


\begin{theorem}\label{th:disjDB}
Given a disjunctive database $\DD$, the set $\Set$ of all disjunctive databases having the same minimal
models as $\DD$ is equal to:
{
\[
\begin{array}{ll}
$\Set$=\{\DD'\ | & \reduced(\DD) \subseteq \DD'\ \wedge \\
          & \forall d' \in \DD'-\reduced(\DD)\ \exists d \in \reduced(\DD) \mbox{ which subsumes } d'\}
\end{array}
\]
}
\em
\noindent \textbf{Proof.}
We denote by $\SMM(\DD)$ the set of all the disjunctive databases whose
minimal models are $\MM{\DD}$.
In order to prove that $\Set=\SMM(\DD)$, first we show that (1) each disjunctive database in $\Set$ is also in $\SMM(\DD)$
and next that (2) each disjunctive database in $\SMM(\DD)$ is in $\Set$ too.\\
(1) Consider a disjunctive database $\DD'$ in $\Set$.
It is easy to see that $\reduced(\DD') = \reduced(\DD)$.
As a disjunctive database and its reduction have the same minimal models,
$\MM{\DD'} = \MM{\DD}$ and hence $\DD'$ is in $\SMM(\DD)$.\\

\noindent
(2) We show that any disjunctive database not belonging to $\Set$ is not in $\SMM(\DD)$.
We recall that for a disjunction $d$, $S_d$ denotes the set of facts appearing in $d$.
Consider a disjunctive database $\DD_{out}$ which is not in $\Set$.
Two cases may occur: (a) $\reduced(\DD) \not\subseteq \DD_{out}$ or
(b) $\reduced(\DD) \subseteq \DD_{out}$ and $\exists d' \in \DD_{out}-\reduced(\DD)$ s.t. there is no $d \in \reduced(\DD)$ which subsumes $d'$.\\
(a) As $\reduced(\DD) \not\subseteq \DD_{out}$, there is a disjunction $a$
in $\reduced(\DD)$ which is not in $\DD_{out}$.
Two cases may occur:
\begin{itemize}
\item
there exists $a_1 \in \DD_{out}$ which subsumes $a$;
\item
the previous condition does not hold.
\end{itemize}
Let us consider the first case and let $I$ be the interpretation $S - S_{a_1}$
where $S$ is the set of facts appearing in $\reduced(\DD)$.
It is easy to see that $I$ is a model of $\reduced(\DD)$
(the only disjunctions that $I$ could not satisfy are those ones that contain
only facts in $S_{a_1}$; such disjunctions are not in
$\reduced(\DD)$ as they  subsume $a$ and
$\reduced(\DD)$ does not contain two disjunctions s.t.
one subsumes the other).
Then, there exists $M \subseteq I$ which is a minimal model of $\reduced(\DD)$.
As $a_1 \in \DD_{out}$, each model of $\DD_{out}$ contains a fact in $S_{a_1}$,
then $M$ is not a minimal model of $\DD_{out}$ and so $\MM{\reduced(\DD)} \neq \MM{\DD_{out}}$.
Hence $\DD_{out} \not\in \SMM(\DD)$.\\
We consider now the second case.
We show that $\DD_{out} \not\in \SMM(\DD)$ in a similar way to the previous case.
Let $I$ be the interpretation $S - S_{a}$
where $S$ is the set of facts appearing in $\DD_{out}$.
It is easy to see that $I$ is a model of $\DD_{out}$
(the only disjunctions that $I$ could not satisfy are those ones which contain
only facts in $S_{a}$; such disjunctions are not in
$\DD_{out}$ as $\DD_{out}$ contains neither $a$ nor a disjunction which subsumes $a$).
Then, there exists $M \subseteq I$ which is a minimal model of $\DD_{out}$.
As $a \in \reduced(\DD)$, each model of $\reduced(\DD)$ contains a fact in $S_{a}$,
then $M$ is not a minimal model of $\reduced(\DD)$; hence $\DD_{out} \not\in \SMM(\DD)$.\\
(b) Let $I$ be the interpretation $S - S_{d'}$
where $S$ is the set of facts appearing in $\reduced(\DD)$.
It is easy to see that $I$ is a model of $\reduced(\DD)$
(the only disjunctions that $I$ could not satisfy are $d'$ and those ones which subsume $d'$).
Then, there exists $M \subseteq I$ which is a minimal model of $\reduced(\DD)$.
As $d' \in \DD_{out}$, each model of $\DD_{out}$ contains a fact in $S_{d'}$,
then $M$ is not a minimal model of $\DD_{out}$; hence $\DD_{out} \not\in \SMM(\DD)$.~\hfill$\Box$
\end{theorem}

\begin{corollary}\label{cor:algorithm-smallest}
Given a database $\D$ and a set $\F$ of denial constraints,
then \Dfr\ is the canonical disjunctive database whose minimal models are the
repairs for $\D$ and $\F$.\\

\em
\noindent \textbf{Proof.}
Straightforward from Theorem~\ref{th:algo_correctness}~and~\ref{th:disjDB}.\hfill$\Box$
\end{corollary}

From now on, we will denote by $\Dmin{\D}{\F}$  the canonical disjunctive database
whose minimal models are the repairs for a database $\D$ and
a set $\F$ of denial constraints.
Whenever $\D$ and $\F$ are clear from the context, we simply write $\Dmi$ instead of $\Dmin{\D}{\F}$.

\section{Functional dependencies}\label{sec:specialcases}

In this section we study the size of the canonical disjunctive database representing
the repairs of a database in the presence of functional dependencies.
Specifically, we show that when the constraints consist of only one key, the canonical disjunctive database
is of linear size, whereas for one non-key functional dependency or two keys the
size of the canonical database may be exponential.\\
We observe that in the presence of only one functional dependency,
the conflict hypergraph has a regular structure that ``induces'' a regular
disjunctive database which can be identified without performing Algorithm~1.
When two key constraints are considered, we are not able to provide such a characterization;
this is because the conflict hypergraph can have an irregular structure and it is
harder to identify a pattern for \Dmi.

Given a  disjunction $d$, we denote by $||d||$ the number of facts occurring in $d$.
The size of a disjunctive database $\DD$, denoted as $||\DD||$, is the number of facts occurring in it,
that is $||\DD||=\sum_{d \in \mathcal{D}} ||d||$.
We study the size $||\Dmi||$ of $\Dmi$ as a function of the size of the given  database.
\\

\noindent
{\em\textbf{One key}.}
Given a relation $r$ and a key constraint $k$ stating that the set $X$ of attributes is a key of $r$,
we denote by $\cliques{r}{k}$ the partition of $r$
into $n = |\pi_{_{X}}(r)|$ sets $\Cl_1, \dots, \Cl_n$, called \emph{cliques}, s.t. each $\Cl_i$
does not contain two facts with different values on $X$.
Observe that
(i) facts in the same clique are pairwise conflicting with each other,
(ii) the set of repairs of $r$ w.r.t. $k$ is
$\{ \{t_1, \dots, t_n\}\ |\ t_i \in \Cl_i \mbox{ for } i=1..n\}$.

\begin{proposition}
Given a relation $r$ and a key constraint $k$,
then \Dmi\ is equal to
\[
\{t_1 \vee \dots \vee t_m\ |\ \exists C = \{t_1,\dots, t_m\}\in \cliques{r}{k}\}
\]

\em
\noindent \textbf{Proof.}
It is straightforward to see that the minimal models of the disjunctive database reported above
are the repairs of $r$ w.r.t. $k$;
since it coincides with its reduction, Theorem~\ref{th:disjDB} implies
that it is the canonical one.~\hfill$\Box$
\end{proposition}

It is easy to see that when one key constraint is considered, $||\Dmi||=|r|$.

\begin{proposition}
Given a relation and a key constraint,
$\Dmi$ is computed in polynomial time by Algorithm~1.\\

\em
\noindent \textbf{Proof.}
It is easy to see that after the first loop (steps~3-5) Algorithm~1 produces \Dmi\ and,
after that, step~9 is never performed.~\hfill$\Box$
\end{proposition}

%
%
%

\noindent
{\em\textbf{Two keys}.}
We now show that, in the presence of two key constraints, \Dmi\ may have exponential size.
Let $\D_n\ (n>0)$ be the family of databases, containing $3n$ facts, of the following form:

\begin{center}
\begin{tabular}{c|c|c|}
\cline{2-3}
             &  $A$      &   $B$\\
\cline{2-3}
$t_{11}\ $   &  $a$      &   $b_1$\\
$\vdots\ $   &  $\vdots$ &   $\vdots$\\
$t_{n1}\ $   &  $a$      &   $b_n$\\
$t_{12}\ $   &  $a_1$    &   $b_1$\\
$t_{13}\ $   &  $a_1$    &   $b_1'$\\
$\vdots$     &  $\vdots$ &   $\vdots$\\
$t_{n2}\ $   &  $a_n$    &   $b_n$\\
$t_{n3}\ $   &  $a_n$    &   $b_n'$\\
\cline{2-3}
\end{tabular}
\end{center}

\noindent\\
Let $\D \in \D_n$ and $A,B$ be two keys.
The conflict hypergraph for $D$ w.r.t. the two key constraints consists of the following edges:
\[
\{\{t_{i1}, t_{j1}\}\ |\ 1 \leq i,j \leq n\  \wedge\ i\neq j\}\ \cup\ \{\{t_{i1}, t_{i2}\}\ |\ 1 \leq i \leq n \}\ \cup\ \{\{t_{i2}, t_{i3}\}\ |\ 1 \leq i \leq n \}
\]
\noindent
Thus, the conflict hypergraph contains a clique $\{t_{11}, \dots,t_{n1}\}$ of size $n$ and, moreover,
$t_{i1}$ is connected to $t_{i2}$ which is in turn connected to $t_{i3}$ ($i=1..n$).

\begin{example}
The conflict hypergraph for a database in $D_4$, assuming that $A$ and $B$ are two keys, is reported in Figure~1.

\begin{figure}[h]\label{fig:conflictgraphD4}
    \centering
    \includegraphics[width=45mm]{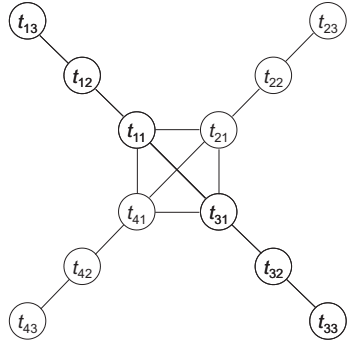}
    \caption{Conflict hypergraph for a database in $\D_4$ w.r.t. $A,B$ key constraints}
\end{figure}
\end{example}

The following proposition identifies the canonical disjunctive database for a database in $\D_n$
for which $A$ and $B$ are keys; such a disjunctive database has exponential size.

\begin{proposition}\label{pro:2keys-size}
Consider a database $\D$ in $\D_n$ and a set of constraints $\F$ consisting of two keys, $A$ and $B$.
Then \Dmi\ is equal to $\DD$ where
\[
\DD=\{t_{i2} \vee t_{i3}\ |\ 1 \leq i \leq n  \}\ \cup\\
\{t_{i1} \vee t_{i2} \vee \bigvee_{z=1..n\ \wedge\ z \neq i} t_z \ |\ 1 \leq i \leq n\ \wedge t_z \in \{t_{z1}, t_{z3}\} \}
\]

\em
\noindent
\textbf{Proof.}
First of all, we show that the minimal models of $\DD$ are the repairs of $\D$ w.r.t. $\F$;
in particular we prove that (1) $\MM{\DD} \subseteq \rep{\D}{\F}$ and (2) $\MM{\DD} \supseteq \rep{\D}{\F}$.\\
(1) Consider a minimal model $m \in \MM{\D}$.
First we show that (a) $m$ is consistent w.r.t. $\F$ and next (b) that it is maximal.\\
(a) Let $E$ be the set of edges of $\mathcal{G}_{D,F}$.
First we show that for each $e = \{t',t''\}$ in $E$ and pair of disjunctions $d' = t' \vee D'$,
$d'' = t'' \vee D''$ in $\DD$ s.t. $D'$ (resp. $D''$) does not contain $t''$ (resp. $t'$), there is
a disjunction in $\DD$ which is equal to or subsumes $D' \vee D''$;
next we show that this property implies that $m$ is consistent w.r.t. $F$.
We recall that $E$ is the union of the following three sets:
\[
E_1 = \{\{t_{i1}, t_{j1}\}\ |\ 1 \leq i,j \leq n\  \wedge i\neq j\}
\]
\[
E_2 = \{\{t_{i1}, t_{i2}\}\ |\ 1 \leq i \leq n \}
\]
\[
E_3 = \{\{t_{i2}, t_{i3}\}\ |\ 1 \leq i \leq n \}
\]
Let us consider the case where $e \in E_1$, that is
$e=\{t_{i1}, t_{j1}\}\ (1 \leq i,j\leq n\ \wedge\ i\neq j)$.
Then a disjunction in $\DD$ containing $t_{i1}$ but not $t_{j1}$ is of the form
\[
d'_1\ :\ t_{i1} \vee t_{i2} \vee t_{j3} \vee \bigvee_{z=1..n\ \wedge\ z\neq i,j} t_z'
\]
where $t_z' \in \{t_{z1}, t_{z3}\}$, or of the form
\[
d'_2\ :\ t_{h1} \vee t_{h2} \vee t_{i1} \vee t_{j3} \vee \bigvee_{z=1..n\ \wedge\ z\neq h,i,j} t_z'
\]
where $1 \leq h \leq n\ \wedge\ h\neq i,j$ and $t_z' \in \{t_{z1}, t_{z3}\}$.
Likewise, a disjunction in $\DD$ that contains $t_{j1}$ but not $t_{i1}$ is of the form
\[
d''_1\ :\ t_{j1} \vee t_{j2} \vee t_{i3} \vee \bigvee_{z=1..n\ \wedge\ z\neq i,j} t_z''
\]
where $t_z'' \in \{t_{z1}, t_{z3}\}$, or of the form
\[
d''_2\ :\ t_{k1} \vee t_{k2} \vee t_{j1} \vee t_{i3} \vee \bigvee_{z=1..n\ \wedge\ z\neq k,i,j} t_z''
\]
where $1 \leq k \leq n\ \wedge\ k\neq i,j$ and $t_z'' \in \{t_{z1}, t_{z3}\}$.
In all the four possible cases, there is disjunction in $\DD$ which subsumes $D' \vee D''$:
\begin{itemize}
\item
if $d' = d'_1$ and $d'' = d''_1$, then there exist both $t_{j2} \vee t_{j3}$ and $t_{i2} \vee t_{i3}$ in $\DD$
which subsume $D' \vee D''$;
\item
if $d' = d'_1$ and $d'' = d''_2$, then there exists $t_{i2} \vee t_{i3}$ in $\DD$ which subsumes $D' \vee D''$;
\item
if $d' = d'_2$ and $d'' = d''_1$, then there exists $t_{j2} \vee t_{j3}$ in $\DD$ which subsumes $D' \vee D''$;
\item
if $d' = d'_2$ and $d'' = d''_2$, then both
$t_{h1} \vee t_{h2} \vee t_{i3} \vee t_{j_3} \vee \bigvee_{z=1..n\ \wedge\ z\neq h,i,j} t_z'$
and $t_{k1} \vee t_{k2} \vee t_{i3} \vee t_{j_3} \vee \bigvee_{z=1..n\ \wedge\ z\neq k,i,j} t_z''$,
which are in $\DD$, subsume $D' \vee D''$.
\end{itemize}
\noindent
Let us consider the case where $e \in E_2$, namely $e = \{t_{i1}, t_{i2}\}\ (1 \leq i \leq n)$.
A disjunction containing $t_{i1}$ but not $t_{i2}$ is of the form
\[t_{k1} \vee t_{k2} \vee t_{i1} \vee \bigvee_{z=1..n\ \wedge\ z\neq i,k} t_z\]
where $1 \leq k \leq n\ \wedge\ k\neq i$ and $t_z \in \{t_{z1}, t_{z3}\}$, whereas a disjunction containing $t_{i2}$ but not $t_{i1}$
is of the form $t_{i2} \vee t_{i3}$.
Thus, $D' \vee D''$, which is equal to
\[
t_{k1} \vee t_{k2} \vee t_{i3} \vee \bigvee_{z=1..n\ \wedge\ z\neq i,k} t_z\
\]
is in $\DD$.
Finally, consider the last case where $e \in E_3$, that is $e = \{t_{i2}, t_{i3}\}\ (1 \leq i \leq n)$.
A disjunction containing $t_{i2}$ but not $t_{i3}$ is of the form
\[
t_{i1} \vee t_{i2} \vee \bigvee_{z=1..n\ \wedge\ z \neq i} t'_z
\]
where $t'_z \in \{t_{z1}, t_{z3}\}$,
whereas a disjunction containing $t_{i3}$ but not $t_{i2}$ is of the form
\[
t_{h1} \vee t_{h2} \vee t_{i3} \vee \bigvee_{z=1..n\ \wedge\ z \neq h,i} t''_z
\]
where $1 \leq h \leq n\ \wedge\ h \neq i$ and $t''_z \in \{t_{z1}, t_{z3}\}$;
$D' \vee D''$ is subsumed or equal to the disjunction
\[
t_{h1} \vee t_{h2} \vee t_{i1} \vee \bigvee_{z=1..n\ \wedge\ z \neq h,i} t''_z
\]
which is in $\DD$.

Assume by contradiction that $m$ is not consistent.
Then there are two facts $t_a, t_b \in m$ s.t. $\{t_a,t_b\}\in E$.
Let $S_{t_a} = \{D\ |\ t_a \vee D \in \DD \mbox{ and }D \mbox{ does not contain }t_b\}$
and $S_{t_b} = \{D\ |\ t_b \vee D \in \DD \mbox{ and }D \mbox{ does not contain }t_a\}$.
As we have seen before, both these sets are not empty.
We have previously proved that for each $D_a\in S_{t_a}$ and $D_b\in S_{t_b}$ there is
a disjunction in $\DD$ which equals or subsumes $D_a \vee D_b$.
Then, there is a set $S_{t_x}$ among $S_{t_a}$ and $S_{t_b}$ s.t. $m$ satisfies each $D$ in $S_{t_x}$,
otherwise there would be $D_a\in S_{t_a}, D_b\in S_{t_b}$ and a disjunction in $\DD$
which is equal to or subsumes $D_a \vee D_b$ which is not satisfied by $m$.
Consider the interpretation $m' = m - \{t_x\}$ and let $t_y$ be the fact among $t_a$ and $t_b$ which is not $t_x$.
We now show that $m'$ is a model, that contradicts the minimality of $m$.
Clearly, $m'$ satisfies every disjunction in $\DD$ which does not contain $t_x$.
As for the disjunctions in $\DD$ containing $t_x$, it is easy to see that they are satisfied by $m'$:
disjunctions containing $t_y$ are satisfied since $t_y \in m'$,
disjunctions not containing $t_y$ are satisfied as well since $m'$ satisfies every disjunction in $S_{t_x}$.
Hence $m$ is consistent w.r.t. $\F$.\\
(b) Now we prove that $m$ is a maximal (consistent) subset of $\D$.
First of all, we note that for each fact $t \in \D$ there is a disjunction $t \vee t_1 \vee \dots \vee t_n$
in $\DD$ s.t. $t_1, \dots, t_n$ are facts conflicting with $t$:
\begin{itemize}
\item
for the facts $t_{i2}$ and $t_{i3}$ ($i=1..n$) such disjunctions are $t_{i2} \vee t_{i3}$;
\item
for the facts $t_{i1}$ ($i=1..n$) such disjunctions are
$t_{i1} \vee t_{i2} \vee \bigvee_{z=1..n\ \wedge\ z\neq i} t_{z1}$.
\end{itemize}
Assume by contradiction that $m$ is not a maximal (consistent) subset of $\D$.
Then there exists $m' \supset m$ which is consistent.
Let $t$ be a fact in $m'$ but not in $m$.
Since $m'$ is consistent, each fact conflicting with $t$ is not in $m'$ and, thus, neither in $m$.
This implies that $m$ doesn't satisfy the disjunction $t \vee t_1 \vee \dots \vee t_n$
containing $t$ and some fact conflicting with it: the fact that $m$ is a model is contradicted.

\noindent
(2) Consider a repair $r$ for $\D$ and $\F$.
We show first (a) that $r$ is a model of $\DD$ and next (b) that it is a minimal model.\\
(a) Suppose by contradiction that $r$ is not a model of $\DD$, then there is a disjunction $d \in \DD$
which is not satisfied by $r$.
Specifically, $d$ is either of the form  $t_{i2} \vee t_{i3}\ (1\leq i \leq n)$ or
$t_{i1} \vee t_{i2} \vee \bigvee_{z=1..n\ \wedge\ z \neq i} t_z$,  $1\leq i \leq n$ and $t_z \in \{t_{z1}, t_{z3}\}$.
In the former case, $r \cup \{t_{i3}\}$ is consistent, since the only fact conflicting with $t_{i3}$,
namely $t_{i2}$, is not in $r$.
This contradicts the maximality of $r$.
As for the latter case, let $T_3 = \{t_{j3}\ |\ t_{j3} \mbox{ appears in }d\}$.
For each $t_{j3} \in T_3$ we have that $t_{j2}\in r$, because $r$ does not contain $t_{j3}$ and $t_{j3}$ is
conflicting only with $t_{j2}$ (if $t_{j2}$ was not in $r$, then $r$ would not be maximal).
Then for each $t_{j3} \in T_3$, since $r$ contains $t_{j2}$, it does not contain $t_{j1}$
otherwise it would not be consistent.
Thus $r$ does not contain any fact $t_{k1}$ with $1\leq k \leq n\ \wedge\ k\neq i$.
Since $r$ contains neither the facts $t_{k1}$'s nor $t_{i2}$,
which are all the facts conflicting with $t_{i1}$,
then $r \cup \{t_{i1}\}$ is consistent (observe that $t_{i1} \not\in r$).
This contradicts the maximality of $r$.
Hence $r$ is a model of $\DD$.\\
(b) We now show that $r$ is a minimal model of $\DD$ reasoning by contradiction.
Assume that there exists a model $m' \subset r$ of $\DD$ and let $t$ be a fact in $r$ but not in $m'$.
All the facts conflicting with $t$ are not in $r$ as $r$ is consistent.
The same holds for $m'$ since it is a (proper) subset of $r$.
We recall that for each fact $t' \in \D$ there is a disjunction
in $\DD$ containing $t'$ and only facts conflicting with $t'$;
then there is a disjunction $d: t \vee t_1 \vee \dots \vee t_n$
in $\DD$ s.t. $t_1, \dots, t_n$ are facts conflicting with $t$.
Since $m'$ does not satisfy $d$, it is not a model, thus we get a contradiction.
Hence $r$ is a minimal model of $\DD$.\\

\noindent
We have shown that the minimal models of $\DD$ are the repairs of $\D$ w.r.t. $\F$.
Since $\DD = \reduced(\DD)$, from Theorem~\ref{th:disjDB} we have that $\DD$ is the canonical
disjunctive database whose minimal models are the repairs of $\D$ w.r.t. $\F$.~\hfill$\Box$
\end{proposition}

\begin{corollary}
Consider a database $\D$ in $\D_n$ and let $A$ and $B$ be two keys;
$||\Dmi|| =  2n+(n+1)\cdot n2^{n-1}$.\\

\em
\noindent
\textbf{Proof.}
From Proposition~\ref{pro:2keys-size}, it is easy to see that \Dmi\ contains
$n$ disjunctions of 2 facts and $n2^{n-1}$ disjunctions of $n+1$ facts.\hfill$\Box$
\end{corollary}

\noindent
{\em\textbf{One functional dependency}.}
Given a relation $r$ and a functional dependency $f: X \rightarrow Y$,
we denote by $\cliques{r}{f}$ the partition of $r$
into $n = |\pi_{_{X}}(r)|$ sets $\Cl_1, \dots, \Cl_n$, called \emph{cliques}, s.t. each $\Cl_i$
does not contain two facts with different values on $X$.
For each clique $\Cl_i$ in $\cliques{r}{f}$ we denote by $\cls(\Cl_i)$ the partition of $\Cl_i$ into
$m_i = |\pi_{_{Y}}(\Cl_i)|$ sets $\Gr_1, \dots, \Gr_{m_i}$, called \emph{clusters},
s.t. each cluster doesn't contain two facts with different values on $Y$.
It is worth noting that
(i) facts in the same cluster are not conflicting each other,
(ii) given two different clusters $\Gr_1$, $\Gr_2$ of the same clique,
each fact in $\Gr_1$ (resp. $\Gr_2$) is conflicting with every fact in $\Gr_2$ (resp. $\Gr_1$),
(iii) the set of repairs of $r$ w.r.t. $f$ is
$\{\Gr_1 \cup \dots \cup \Gr_n\ |\ \Gr_i \in \cls(\Cl_i) \mbox{ for } i=1..n\}$.

\begin{proposition}
Given a relation $r$ and a functional dependency $f$, then $\Dmi$ is equal to $\DD$ where
\[
\begin{array}{ll}
\DD=\{t_1 \vee \dots \vee t_k\ | & \exists \Cl \in \cliques{r}{f}\ s.t.\ \cls(\Cl)=\{\Gr_1, \dots, \Gr_k\}\\
                                           & \mbox{ and } t_1 \in \Gr_1, \dots, t_k \in \Gr_k \}
\end{array}
\]

\em
\noindent \textbf{Proof.}
We show first (1) that each minimal model of $\DD$ is a repair for $r$ and $f$
and next (2) that each repair of $r$ w.r.t. $f$ is a minimal model of $\DD$.\\
(1) Consider a minimal model $m$ of $\DD$.
Let $\cliques{r}{f} = \{\Cl_1, \dots, \Cl_n\}$ be the cliques for $r$ and $f$.
For each clique $\Cl_i$ in $\cliques{r}{f}$ there is a cluster $\Gr_j$ in $\cls(\Cl_i)=\{\Gr_1, \dots, \Gr_k\}$
s.t. $\Gr_j \subseteq m$ (otherwise $m$ would not satisfy the disjunction $t_1 \vee \dots \vee t_k$
in $\DD$ where $t_h \in \Gr_h$ and $t_h \not\in m$, $h=1..k$).
Let $\overline{\Gr}_1, \dots, \overline{\Gr}_n$ be such clusters, where each $\overline{\Gr}_l$ is a cluster of
$\Cl_l$ for $l=1..n$.
Since $\overline{\Gr}_1 \cup \dots \cup \overline{\Gr}_n \subseteq m$ and
$\overline{\Gr}_1 \cup \dots \cup \overline{\Gr}_n \models \DD$,
then $m = \overline{\Gr}_1 \cup \dots \cup \overline{\Gr}_n$, which is, as we have observed before, a repair.\\
(2) Consider a repair $s$ in $\rep{r}{f}$.
As $s$ consists of one cluster for each clique, it is easy to see that $s$ is a model of $\DD$.
We show that $s$ is minimal by contradiction assuming that there exists $s' \subset s$
which is a model of $\DD$.
Let $t$ be a fact in $s$ which is not in $s'$.
Let $\Cl_t$ and $\Gr_t$ be the clique and the cluster, respectively, containing $t$;
moreover let $\cls(\Cl_t)=\{\Gr_t, \Gr_1, \dots, \Gr_k\}$.
The disjunction $t \vee t_1 \vee \dots \vee t_k$, where $t_i \in \Gr_i,\ i=1..k$, which is
in $\DD$, is not satisfied by $s'$ as $s'$ contains exactly one cluster per clique
(thus it does not contain any fact in $\Gr_i,\ i=1..k$) and does not contain $t$.
This contradicts the fact that $s'$ is a model.
So $s$ is a minimal model of $\DD$.\\
Hence the minimal models of $\DD$ are exactly the repairs for $r$ and $f$;
as $\DD$ is equal to its reduction, Theorem~\ref{th:disjDB} entails that
$\DD=\Dmi$.~\hfill$\Box$
\end{proposition}

Clearly, the size of \Dmi\ may be exponential if the functional dependency is a non-key dependency, as shown in the following example.

\begin{example}\label{ex:oneFD}
Consider the relation $r$, consisting of $2n$ facts, reported below and the non-key functional dependency $A \rightarrow B$.
\begin{center}
\begin{tabular}{c|c|c|c|}
\cline{2-4}
              &  $A$      &   $B$      &   $C$ \\
\cline{2-4}
$t_1' \ $     &  $a$      &   $b_1$    &  $c_1$\\
$t_1''\ $     &  $a$      &   $b_1$    &  $c_2$\\
$\vdots\ $    &  $\vdots$ &   $\vdots$ &  $\vdots$\\
$t_n' \ $     &  $a$      &   $b_n$    &  $c_1$\\
$t_n''\ $     &  $a$      &   $b_n$    &  $c_2$\\
\cline{2-4}
\end{tabular}
\end{center}

\noindent \\
There is a unique clique consisting of $n$ clusters $\Gr_i = \{t_i', t_i''\}$, $i=1..n$.
Then $\Dmi = \{ t_1 \vee \dots \vee t_n\ |\ t_i \in \Gr_i \mbox{ for } i=1..n\}$
and $||\Dmi|| = n2^n$.
\end{example}

\section{Cardinality-based repairs}\label{sec:card-repairs}

In this section we consider \emph{cardinality-based repairs}, that is consistent databases which
minimally differ from the original database in terms of the number of facts in the symmetric difference
(in the previous sections we have considered consistent databases for which the symmetric difference
is minimal under set inclusion, we will refer to them as \emph{\repair s}).

We show that, likewise to what has been presented in Section~\ref{sec:specialcases}, the size of the canonical
disjunctive database (representing the cardinality-based repairs) is linear when only one key constraint
is considered, whereas it may be exponential when two keys or one non-key functional dependency are considered.

It is easy to see that in the presence of only one key constraint the cardinality-based repairs coincide
with the \repair s, so the canonical disjunctive database is of linear size.

When the constraints consists of one functional dependency, it is easy to see that if for every clique
its clusters have the same cardinality, then the cardinality-based repairs coincide with the \repair s.
This is the case for the database of Example~\ref{ex:oneFD}, where the size of the canonical disjunctive
database is exponential.

Finally, we consider the case where two key constraints are considered.
We directly show that the size of the canonical disjunctive database is also exponential.

\begin{lemma}\label{lem:repairs2keys}
Consider a database $\D$ in $\D_n$ and a set of integrity constraints $\F$ consisting of two keys, $A$ and $B$.
Then the set of \repair s is is equal to $R$ where
\[
R=\{\{t_{12},\dots,t_{n2}\}\} \cup
\{\ \{t_{i1},t_{i3}\}\cup \bigcup_{j=1..n\ \wedge\ j\neq i} \{t_j\}\ |\ 1\leq i \leq n\  \wedge\  t_j \in \{t_{j2},t_{j3}\} \}
\]
\em
\noindent
\textbf{Proof.}
It is easy to see that each database in $R$ is a \repair.\\
Consider a \repair\ $r$ of $\D$ w.r.t. $\F$.
We show that $r$ is in $R$ using reasoning by cases:
\begin{itemize}
\item
Suppose that $t_{13}\in r$.
Then $t_{12}\not\in r$ and either (1) $t_{11}\in r$ or (2) $t_{11}\not\in r$.
\begin{enumerate}
\item
Since $t_{11}\in r$, for $j=2..n$ $t_{j1} \not\in r$ and either $t_{j2}$ or $t_{j3}$ is in $r$,
that is $r=\{t_{11},t_{13},t_{2},\dots,t_{n}\}$ where $t_j \in \{t_{j2},t_{j3}\},\ j=2..n$.
It is easy to see that $r \in R$.
\item
Since $t_{11}\not\in r$, there exists $t_{k1}\in r$ with $2 \leq k \leq n$.
Then $t_{k2} \not\in r$ and $t_{k3} \in r$.
For $j=2..n\ \wedge\ j\neq k$, $t_{j1} \not\in r$ and either $t_{j2}$ or $t_{j3}$ is in $r$,
that is $r=\{t_{13},t_{k1},t_{k3}\}\cup \bigcup_{j=2..n\ \wedge\ j\neq k} \{t_j\}$ where $t_j \in \{t_{j2},t_{j3}\}$.
Clearly, $r \in R$.
\end{enumerate}
\item
Suppose that $t_{13}\not\in r$.
Then $t_{12}\in r$ and $t_{11}\not\in r$.
Two cases may occur: either (1) there exists $t_{k1}\in r$ with $2 \leq k \leq n$ or (2) $t_{j1}\not\in r$
for $j=1..n$.
\begin{enumerate}
\item
Since $t_{k1}\in r$ then $t_{k2}\not\in r$ and $t_{k3}\in r$.
For $j=2..n\ \wedge\ j\neq k$ $t_{j1} \not\in r$ and either $t_{j2}$ or $t_{j3}$ is in $r$,
that is $r=\{t_{12},t_{k1},t_{k3}\}\cup \bigcup_{j=2..n\ \wedge\ j\neq k} \{t_j\}$ where $t_j \in \{t_{j2},t_{j3}\}$.
It is easy to see that $r \in R$.
\item
$r=\{t_{12},\dots,t_{n2}\}$ which is in $R$.~\hfill$\Box$
\end{enumerate}
\end{itemize}
\end{lemma}

\begin{corollary}\label{cor:card-basedRepairs2keys}
Consider a database $\D$ in $\D_n$ and a set of integrity constraints $\F$ consisting of two keys, $A$ and $B$.
Then the set of cardinality-based repairs is
\[
\{\ \{t_{i1},t_{i3}\}\cup \bigcup_{j=1..n\ \wedge\ j\neq i} \{t_j\}\ |\ 1\leq i \leq n\  \wedge\  t_j \in \{t_{j2},t_{j3}\} \}
\]
\em
\noindent
\textbf{Proof.}
Straightforward from Lemma~\ref{lem:repairs2keys}.~\hfill$\Box$
\end{corollary}

The following proposition identifies the canonical disjunctive database for a database in $\D_n$
for which $A$ and $B$ are keys; such a disjunctive database is of exponential size.
In the following proposition and corollary, \Dmi\ denotes the canonical disjunctive database representing
the set of cardinality-based repairs.

\begin{proposition}\label{pro:2keys-size-card}
Consider a database $\D$ in $\D_n$ and a set of integrity constraints $\F$ consisting of two keys, $A$ and $B$.
Then the canonical disjunctive database \Dmi\ is equal to $\DD$ where
\[
\DD=\{t_{i2} \vee t_{i3}\ |\ 1 \leq i \leq n \}\ \cup\\
\{t_1 \vee \dots \vee t_n |\ t_i\in \{t_{i1},t_{i3}\},\ i=1..n \}
\]
\em
\noindent
\textbf{Proof.}
We first show that (1) each cardinality-based repair of $\D$ w.r.t. $\F$ is a minimal model of $\DD$ and
next that (2) each minimal model of $\DD$ is a cardinality-based repair.

\noindent
(1) Consider a cardinality-based repair $r$ of $\D$ w.r.t. $\F$.
We show first that (a) $r$ is a model of $\DD$ and next that (b) it is  a minimal model.\\
(a) From Corollary~\ref{cor:card-basedRepairs2keys}, it is easy to see that $r$ satisfies
each disjunction $t_{i2} \vee t_{i3}$ in $\DD$, $1 \leq i \leq n$.
Since Corollary~\ref{cor:card-basedRepairs2keys} entails that there exists $1 \leq j \leq n$ s.t. $\{t_{j1},t_{j3}\} \subseteq r$,
then $r$ satisfies each disjunction $t_1 \vee \dots \vee t_n$ in $\DD$ (where $t_i\in \{t_{i1},t_{i3}\},\ i=1..n$).
Thus $r$ is a model of $\DD$.\\
(b) We observe that for each fact $t \in \D$ there is a disjunction $t \vee t_1 \vee \dots \vee t_n$
in $\DD$ s.t. $t_1, \dots, t_n$ are facts conflicting with $t$: for the facts $t_{i2}$ and $t_{i3}$ $(i=1..n)$
such disjunctions are $t_{i2} \vee t_{i3}$; for the facts $t_{i1}$ $(i=1..n)$ there is the disjunction
$t_{11}\vee \dots \vee t_{n1}$.
In the same way as in Proposition~\ref{pro:2keys-size}, it can be shown that $r$ is a minimal model of $\DD$.

\noindent
(2) Consider a minimal model $m$ of $\DD$.
The fact that $m$ is a \repair\ of $\D$ w.r.t. $\F$ can be shown in the same way as
in Proposition~\ref{pro:2keys-size}.

\noindent
It is easy to see that $\{t_{12},\dots,t_{n2}\}$ is not a model of $\DD$ and then, from
Lemma~\ref{lem:repairs2keys} and Corollary~\ref{cor:card-basedRepairs2keys}, $m$ is a cardinality-based
repair of $\D$ w.r.t. $\F$.\\

\noindent
We have shown that $\DD$ represents the cardinality-based repairs of $\D$ w.r.t. $\F$;
since $\DD = \reduced(\DD)$, from Theorem~\ref{th:disjDB} we have that $\DD$ is the canonical one.~\hfill$\Box$
\end{proposition}

\begin{corollary}
Consider a database $\D$ in $\D_n$ and let $A$ and $B$ be two keys;
$||\Dmi|| =  2n+n2^n$.\\

\em
\noindent
\textbf{Proof.}
From Proposition~\ref{pro:2keys-size-card}, it is easy to see that \Dmi\ contains $n$ disjunctions of 2 facts
and $2^n$ disjunctions of $n$ facts.\hfill$\Box$
\end{corollary}

\section{Conclusions}\label{sec:conclusions}

In this paper we have addressed the problem of representing, by means of a disjunctive database, the set of repairs
of a database w.r.t. a set of denial constraints.
We have shown that, given a database and a set of denial constraints,
there exists a unique canonical disjunctive database representing their repairs:
any disjunctive database with the same set of minimal models
is a superset of the canonical one, containing in addition disjunctions which are subsumed by the disjunctions
in the canonical one.
We have proposed an algorithm to compute the canonical disjunctive database.
We have shown that the size of the canonical disjunctive database is linear when only one key is considered,
but it may be exponential in the presence of two keys  or one non-key functional dependency.
We have shown that these results hold also when cardinality-based repairs are considered.

Future work in this area could explore different representations for the set of repairs.
For instance, one can consider formulas with negation or non-clausal formulas. Such formulas
can be more succinct than disjunctive databases, making query evaluation, however, potentially harder.
We also observe that in the case of the repairs of a single
relation the resulting disjunctive database consists of disjunctions of elements of this relation.
It has been recognized that such disjunctions should be supported by database management systems
\cite{BSHTW08}.
Moreover, one could consider {\em restricting} inconsistent databases in such a way that the resulting repairs can be
represented by relational databases with {\em OR-objects} \cite{ImNaVa91}.
In this case, one could use the techniques for computing {\em certain} query answers over databases with OR-objects \cite{ImVMVa95}
to compute {\em consistent} query answers over inconsistent databases.
Finally, other kinds of representations of sets of possible worlds, e.g., \emph{world-set decompositions} \cite{AnKoOl07}, should be considered.
For example, the set of repairs of the database in Example \ref{ex:oneFD} can be represented as a world-set
decomposition of polynomial size.

\begin{acknowledgements}
Jan Chomicki acknowledges the support of National Science Foundation under grant IIS-0119186.
Jerzy Marcinkowski was partially supported by Polish Ministry of Science and Higher Education research project N206 022 31/3660, 2006/2009.
\end{acknowledgements}


\bibliographystyle{plain}
\bibliography{bib}

\end{document}